\documentclass[aip,twocolumn,amsmath,amssymb,seceqn,10pt]{revtex4-1}
\usepackage{graphicx}
\usepackage{graphics}
\usepackage{epsfig}
\usepackage{dcolumn}
\usepackage{bm}
\usepackage{makeidx}
\usepackage{subfigure}
\usepackage{color}
\usepackage{longtable}
\usepackage{upgreek}
 \usepackage{multirow}
  \usepackage[table,xcdraw]{xcolor}

\begin{document}

\title{Anomalous Hall antiferromagnets}
\author{Libor \v{S}mejkal}
\affiliation{Institut f\"ur Physik, Johannes Gutenberg Universit\"at Mainz, 55128 Mainz, Germany}
\affiliation{Institute of Physics, Czech Academy of Sciences, Cukrovarnick\'a 10, 162 00, Praha 6, Czech Republic}
\author{Allan H. MacDonald}
\affiliation{Department of Physics, The University of Texas at Austin, Austin, TX 78712, USA}
\author{Jairo~Sinova}
\affiliation{Institut f\"ur Physik, Johannes Gutenberg Universit\"at Mainz, 55128 Mainz, Germany}
\affiliation{Institute of Physics, Czech Academy of Sciences, Cukrovarnick\'a 10, 162 00, Praha 6, Czech Republic}
\author{Satoru Nakatsuji}
\affiliation{Department of Physics, University of Tokyo, Hongo, Bunkyo-ku, Tokyo 113-0033, Japan}
\affiliation{Institute for Solid State Physics, University of Tokyo, Kashiwa, Chiba 277-8581, Japan}
\affiliation{Trans-scale Quantum Science Institute, University of Tokyo, Bunkyo-ku, Tokyo 113-0033, Japan}
\affiliation{Institute for Quantum Matter and Department of Physics and Astronomy, Johns Hopkins University, Baltimore, Maryland 21218, USA}
\author{Tomas~Jungwirth}
\affiliation{Institute of Physics, Czech Academy of Sciences, Cukrovarnick\'a 10, 162 00, Praha 6, Czech Republic}
\affiliation{School of Physics and Astronomy, University of Nottingham, NG7 2RD, Nottingham, United Kingdom}
\date{\today}

\begin{abstract}

The Hall effect, in which current flows perpendicular to an applied electrical bias, has played a prominent role in modern condensed matter physics over much of the subject's history.
Appearing variously in classical, relativistic and quantum guises, it has among other roles contributed
to establishing the band theory of solids, to research on new phases of strongly interacting electrons,
and to the phenomenology of topological condensed matter.
The dissipationless Hall current 
requires  time-reversal symmetry breaking. For over a century it has either been ascribed to externally applied magnetic field and referred to as the ordinary Hall effect, 
or ascribed to spontaneous non-zero total internal magnetization (ferromagnetism) and referred to as the anomalous Hall effect.  It has not commonly been associated with antiferromagnetic order. 
More recently, however, theoretical  predictions and experimental observations have identified large Hall effects in some compensated magnetic crystals, governed by neither of the global
magnetic-dipole symmetry breaking mechanisms mentioned above. The goals of this article are to systematically organize  the present understanding of anomalous antiferromagnetic 
materials that generate a non-zero Hall effect, which we will call anomalous Hall antiferromagnets, and to discuss this class of materials 
in a broader fundamental and applied research context. Our motivation in drawing attention to anomalous Hall antiferromagnets is two-fold. 
First, since Hall effects that are not governed by magnetic dipole symmetry breaking
are at odds with the traditional understanding of the  phenomenon, the topic deserves attention on its own. Second, this new reincarnation has again placed
the Hall effect in the middle of an emerging field of physics at the intersection of multipole magnetism, topological  condensed matter, and spintronics.
\end{abstract}

\maketitle

The alignment of  the ferromagnetic needle of 
a compass along the Earth's magnetic field 
direction is an ancient  example of a mechanical magnetic-dipole 
sensor.  Its electrical counterpart was discovered at the end of the 19$^{\rm th}$ century by Edwin Hall. 
In Hall's ``electric compass", a current that flows in a direction transverse to an applied electrical bias is generated in a conductor in the presence of the magnetic field, 
and the sign of the transverse current flips if the direction of the magnetic field is 
reversed. The magnetic field can couple directly to the charge of particles carrying the electrical current via the Lorentz force, in which case the phenomenon is referred to as the ordinary Hall effect. Alternatively, the magnetic field can induce internal magnetization in the conductor, in analogy to alignment of  the magnetized needle of a compass. 
The contribution to the Hall current due to the coupling of charge carriers to the internal magnetization is commonly called the  anomalous Hall effect.\cite{Nagaosa2010} In ferromagnets with a spontaneous magnetization, the Hall current can occur even in the absence of the external magnetic field. 
Hall sensing of both internal magnetization and external magnetic fields has now maintained its role in both basic research and in applications 
for nearly a century and a half. 

In the middle of the 20$^{\rm th}$ century, condensed matter physics research was dominated by the early successes of quantum mechanics 
and the band theory of solids.  Here the ordinary Hall effect, whose sign and magnitude depend on the sign and density of charge carriers, helped to establish the notions of electron and hole transport in conduction and valence bands, and to  give birth to the field of semiconductor physics. 
The anomalous Hall effect contributed in a complementary area by providing a macroscopic probe of spontaneous  time-reversal (${\cal T}$) symmetry breaking in ferromagnets.
It was understood early on\cite{Pugh1953} that the anomalous Hall effect could not be explained by 
the coupling of internal magnetization to the charge of the carriers by the Lorentz force. Instead, relativistic spin-orbit coupling was 
identified as the link between the ${\cal T}$-symmetry breaking, generated by the ferromagnetic order, and the charge Hall transport.\cite{Pugh1953} 
This led to major controversies on whether the subtle spin-orbit coupling effects in particular materials enter most strongly intrinsically via the relativistic band structure of an unperturbed crystal 
or extrinsically via impurity scattering.\cite{Karplus1954,Smit1958,Berger1970}

Progress arrived from an unexpected direction thanks to the discovery of the 
quantum Hall effect in 1980.\cite{Klitzing1980}
The quantum Hall effect occurs in two-dimensional (2D) electron systems whose non-relativistic energy spectrum is broken into discrete Landau levels by a strong magnetic field. This turns the bulk into an insulator while conduction is allowed only through dissipationless one-dimensional edge channels. Our understanding of the quantum Hall effect has benefited 
greatly from the introduction of the concept of band-structure topology, which along side symmetry can be used to catagorize states 
of matter. The band topology concept helped to explain the observed exact quantization of the Hall resistance because of its 
relationship to Chern numbers, topological Berry phase invariants of the Landau level band structure.\cite{Thouless1982,Berry1984a,Prange1987}  
The quantum Hall effect also played a prominent role in discoveries of new many-body phases of strongly correlated electrons.\cite{Prange1987} It provided, e.g., a rare example of a correlated state directly described by its many-body wavefunction. 

The anomalous Hall effect returned to the forefront of research in the early 2000's when, borrowing from quantum Hall theory, it was identified with Berry phases in the 
spin-orbit coupled relativistic band structures of ferromagnets.\cite{Jungwirth2002,Onoda2002}   By the time of the publication of a comprehensive topical review in 2010,\cite{Nagaosa2010} 
the Berry phase anomalous Hall contribution was established as a leading mechanism for the Hall effect in a broad family of metallic ferromagnets. 
In addition, a systematic parsing of intrinsic and scattering-induced contributions was completed, clarifying the role disorder plays in common metallic ferromagnets.\cite{Nagaosa2010}  

The success of the Berry phase picture of the anomalous Hall effect was one of the precursors of the field of topological insulators \cite{Franz2013,Bradlyn2017,Elcoro2020} 
which has provided a more general classification of Bloch band topology.  
In 2013, it led to the first experimental realization \cite{Chang2013} of the 
quantum anomalous Hall effect - a quantum Hall effect due to ferromagnetism and not to external magnetic fields.
In the meantime, the quantum Hall effect was demonstrated in graphene at room temperature but it requires a strong magnetic field.\cite{Novoselov2007} 
The quantum anomalous Hall effect can be observed at zero magnetic field but, so far, has been limited to Kelvin temperatures.\cite{Tokura2019,Deng2020}  
The search for a high-temperature zero-field quantum anomalous Hall effect is emerging as the ultimate destination
of a more than century long journey through condensed matter, navigated by the intriguing relativistic, quantum, and topological Hall compass. 

\begin{figure}[h!]
\hspace*{-0cm}\epsfig{width=.8\columnwidth,angle=0,file=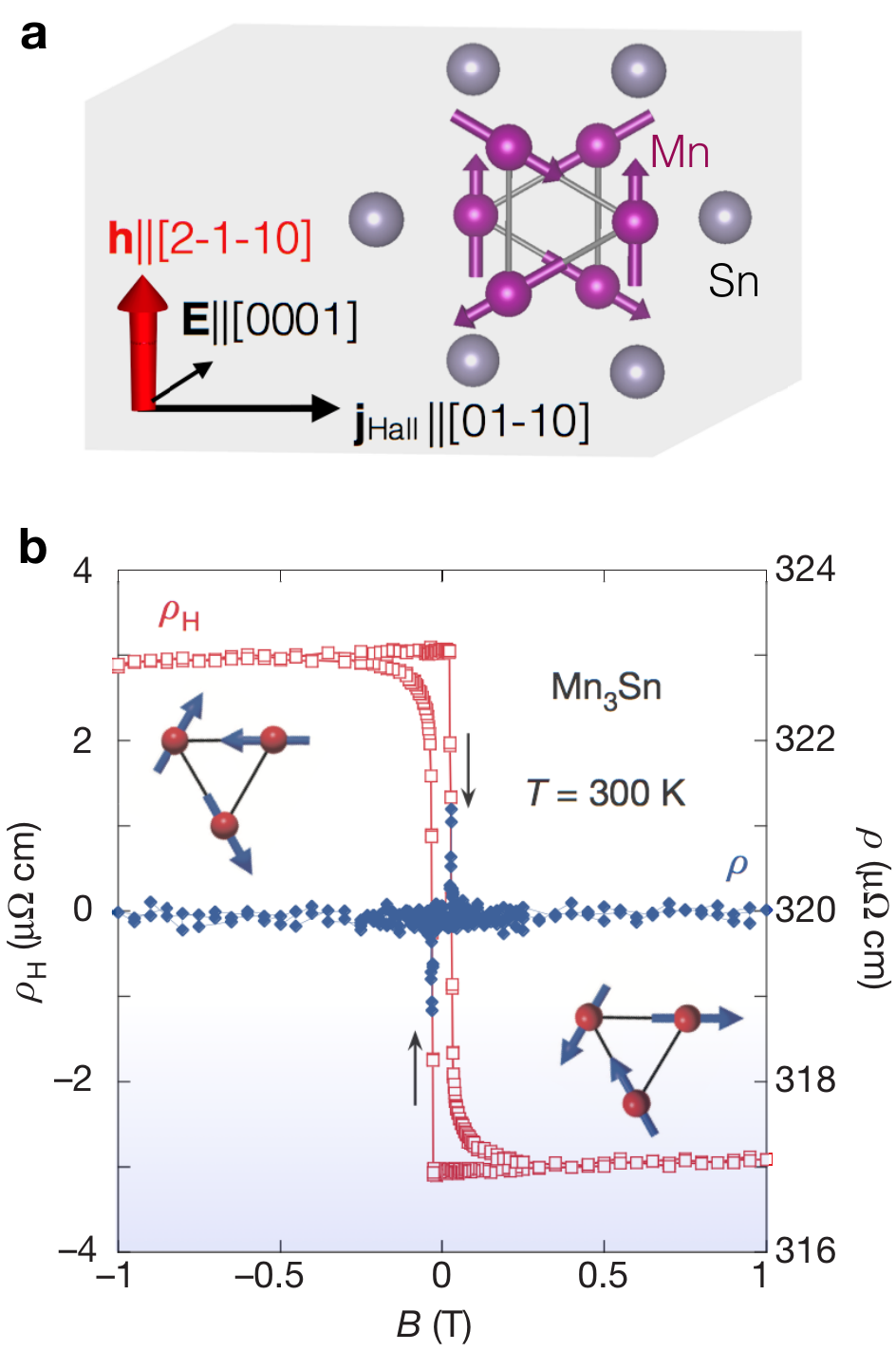}
\caption{{\bf Experimental discovery of the Hall-effect in the compensated non-collinear magnet Mn$_3$Sn.}
{\bf a}, Crystal and magnetic structure of Mn$_3$Sn, and the Hall effect geometry. {\bf b}, Magnetic field dependence of the Hall resistivity (left axis) and the longitudinal resistivity (right axis). 
Strong magnetic fields align magnetic moment in time-reversed states (insets). Panel b is adapted from Ref.~\onlinecite{Nakatsuji2015}.
}
\label{fig1}
\end{figure}

The richness of the physics discovered along the journey, briefly reviewed above, is remarkable given how narrowly the path was bounded. 
Because the Hall effect was viewed for over a century as a magnetic-dipole detector,
its manifestation in materials in which an external magnetic field is absent and internal electron-electron interactions do not generate a net magnetization remained virtually unexplored. 
However, in 2010, a Hall effect was experimentally observed at cryogenic temperatures in the absence of a magnetic dipole  in a spin liquid candidate.\cite{Machida2010}  
Moreover, subsequent studies, guided again by relativistic Berry phase physics,  have identified Hall effects in certain non-collinear antiferromagnets at room temperature 
that are comparable in strength to those of ferromagnets.\cite{Chen2014,Kubler2014,Nakatsuji2015,Kiyohara2015,Nayak2016} 
The discoveries, illustrated in Fig.~\ref{fig1}, force us to abandon the century long view of the Hall effect as a detector of the magnetic dipole.
Hall response is instead often closely related to non-trivial topology in electronic structure which leads to enhanced
response, breaking the paradigm of scaling with total magnetization strength.\cite{Wan2011,Chen2013,Tang2016,Smejkal2016,Kuroda2017,Sakai2018,Liu2018c,Wang2018g,Belopolski2019,Suzuki2019b,Sakai2020,Chen2021a}

Large anomalous Hall effects do not require non-zero total magnetization.
In this review we systematically explain what distinguishes anomalous Hall antiferromagnets from their  
antiferromagnetic cousins which, by symmetry, generate zero Hall effect. We start by returning to the basic concept of ${\cal T}$-symmetry breaking, 
inspecting it in a context beyond the magnetic dipole. This symmetry analysis is presented in the next section. 
Instead of reviewing research progress in chronological order, we take the more pedagogical approach. 
We start from what is perhaps the simplest example of anomalous Hall antiferromagnets,
 the recently discovered crystals containing two collinear antiparallel magnetic sublattices, \cite{Smejkal2020,Feng2020a}
 explaining what makes them special.

We then extend the discussion to multiple sublattices, considering compensated collinear,\cite{Reichlova2020} as well as the non-collinear and non-coplanar magnetic crystals that were historically studied earlier.\cite{Haldane1988,Shindou2001,Tomizawa2009,Tomizawa2010,Machida2010,Ueland2012,Surgers2014,Chen2014,Kubler2014,Nakatsuji2015,Kiyohara2015,Nayak2016} We 
emphasize that the ${\cal T}$-symmetry breaking by the compensated non-coplanar magnetic order can under the right 
circumstances generate a Hall effect even in the absence of the relativistic spin-orbit coupling.\cite{Suzuki2017} 
Note that this non-relativistic mechanism has been referred to as the geometrical or topological Hall effect.\cite{Shindou2001,Machida2010,Ueland2012,Surgers2014}

We use primarily the language of crystal symmetry groups, and also make reference to the magnetic multipole classification.\cite{Suzuki2017,Watanabe2018a,Hayami2018,Suzuki2019a,Hayami2020,Thole2020,Hayami2021,Yatsushiro2021} 
We explain which crystal symmetries favor that non-relativistic electron-electron Coulomb interactions lead to a ${\cal T}$-symmetry broken co-planar magnetic order with a precisely zero magnetization. We show that this  ${\cal T}$-symmetry breaking mechanism, in combination with the relativistic spin-orbit coupling, can for some symmetry types generate the Hall effect. These anomalous Hall antiferromagnets have symmetries which also allow for the existence of a weak relativistic magnetization. We emphasize, however, that it is the above non-relativistic ${\cal T}$-symmetry breaking by the precisely compensated magnetic order, rather than the  weak relativistic magnetization, which governs the Hall effect we focus on in this review.

The symmetry analysis is followed by a section illustrating anomalous Hall  
antiferromagnetism microscopically in terms of band structure and Berry phase calculations in representative crystals. 
We highlight how the symmetry  and microscopic electronic structure of these magnetically compensated crystals 
cooperate to create large Hall signals with a rich phenomenology, and tabulate the anomalous Hall antiferromagnets that have 
been realized experimentally to date.

We conclude the review by discussing the central role of anomalous Hall antiferromagnets in an emerging fundamental and applied field at the
 intersection of multipole magnetism, topological phases, and spintronics. 
The Hall effect, together with electrical readout  by anisotropic magnetoresistance and switching  by the spin-orbit torque,\cite{Shick2010,Park2011b,Marti2014,Zelezny2014,Wadley2016,Jungwirth2016,Wadley2018,Chen2014,Kubler2014,Nakatsuji2015,Kiyohara2015,Nayak2016,Smejkal2020,Feng2020a,Reichlova2020} have been the key phenomena that initiated experimental research in the field of relativistic antiferromagnetic spintronics.\cite{Jungwirth2016,Zelezny2018,Nemec2018,Gomonay2018,Smejkal2017b,Baltz2018,Song2018b,Mizuguchi2019,Siddiqui2020,Fukami2020,Kurenkov2020,Kaspar2021,Ikhlas2017,Higo2018c,Liu2018b,Zelezny2017a,Kimata2019a,Reichlova2019,Tsai2020,Matsuda2020,Samanta2020} 

Recently, studies of anomalous Hall antiferromagnets with collinear order have also drawn the attention to an unexpected alternating spin-polarization in the non-relativistic limit of 
their band structures.\cite{Smejkal2020,Lopez-Moreno2012,Noda2016,Ahn2019,Hayami2019,Yuan2020,Feng2020a,Reichlova2020,Hayami2020,Yuan2021a,Egorov2021,Smejkal2021a,Gonzalez-Hernandez2021}  Instead of regarding these as antiferromagnetic anomalies present only in some materials, the authors of Ref.~\onlinecite{Smejkal2021a} propose viewing them as members of  
 a third magnetic class, altermagnets, along side ferromagnets and antiferromagnets. This class has been systematically delimited in Ref.~\onlinecite{Smejkal2021a} based on a formal non-relativistic spin-group theory.
 From an applied perspective,  a new route has emerged for realizing zero magnetic-dipole analogues of multilayer stacks with non-relativistic giant or tunneling magnetoresistance readout and spin-transfer torque switching.\cite{Reichlova2020,Smejkal2021,Shao2021} These phenomena, based on non-relativistic conserved spin-currents,  underpin commercial ferromagnetic spintronics technologies. \cite{Chappert2007,Ralph2008,Brataas2012,Bhatti2017,Duine2018}

Anomalous Hall antiferromagnets also open a new avenue of research on dissipationless transport and quantum topological phenomena.\cite{Smejkal2017b,Mong2010,Tang2016,Smejkal2016,Felser2017,Kuroda2017,Bradlyn2017,Otrokov2019,Noky2019a,Vergniory2019,Tsai2020,Deng2020,Xu2020,Elcoro2020,Nomoto2020,Tsai2021}  The field can benefit from the rich symmetry landscape of these multipole magnets, and can include
 materials ranging from insulators to superconductors.\cite{Smejkal2021a} Finally, we recall that while our focus in this review is on ordered bulk magnetic crystals, Hall effect mechanisms not governed by a magnetic dipole  were also identified in fluctuating quantum phases, such as chiral spin liquids or in chiral non-collinear structures,\cite{Machida2010,Lux2020} in electrically gated systems including antiferromagnetic Dirac semimetals,\cite{Sivadas2016,Du2020} or  at surfaces of magnetic topological insulators.\cite{Wu2016a,Varnava2018}


\subsection*{Hall effect, time-reversal, and magnetic crystal order}

\noindent{\bf\em Hall vector and time-reversal symmetry breaking.} 
The Hall current ${\bf j}_{\rm Hall}$ corresponds to the antisymmetric part of the conductivity tensor,  $\sigma^{\rm a}_{ij}=-\sigma^{\rm a}_{ji}$, which allows it to be expressed in terms an axial Hall vector ${\bf h}$ as,  ${\bf j}_{\rm Hall}={\bf h}\times{\bf E}$, where ${\bf h} = (\sigma^{\rm a}_{zy}, \sigma^{\rm a}_{xz}, \sigma^{\rm a}_{yx})$.\cite{Landau1984} 
Because Joule heating is given by ${\bf j}\cdot{\bf E}$, the Hall effect is dissipationless.\cite{Landau1984} Therefore, the transformation under ${\cal T}$ of the dissipationless Hall vector must be preserved in the constitutive relation between the Hall current and electric field. Since ${\bf j}_{\rm Hall}$ reverses sign under ${\cal T}$ and ${\bf E}$ does not, ${\bf h}$ must reverse sign under ${\cal T}$, i.e., the Hall vector is a ${\cal T}$-odd axial vector.  According to the Neumann's principle, then, ${\cal T}$-symmetry must be broken in the system for the Hall vector to be nonzero. 
(Note that because the longitudinal current, which generates entropy through Joule heating ${\bf j}\cdot{\bf E}$, is dissipative, the symmetric components of the conductivity tensor are ${\cal T}$-even.\cite{Landau1984})

For the most of the history of the Hall effect research, the considered ${\cal T}$-symmetry breaking mechanisms were either due to an external magnetic field or internal magnetization (ferromagnetism).\cite{Grimmer1993}  The Onsager relations in the presence of the external magnetic field, $\sigma_{ij}({\bf H})=\sigma_{ji}({\bf -H})$, then explicitly  imply that the Hall effect is odd in ${\bf H}$: $\sigma^{\rm a}_{ij}({\bf H})=-\sigma^{\rm a}_{ij}({\bf -H})$ and ${\bf h}({\bf H})=-{\bf h}({\bf -H})$.\cite{Landau1984}
In ferromagnets with a non-zero averaged  internal magnetization ${\bf M}$, the ${\cal T}$-symmetry is broken even at ${\bf H}=0$. 
In analogy to the external magnetic field, the Hall effect is again allowed, with ${\bf h}({\bf M})=-{\bf h}({\bf -M})$.\cite{Shtrikman1965,Grimmer1993} 


In this review we focus on the relationship between ${\cal T}$-symmetry breaking and the Hall effect at ${\bf H}=0$ in 
the case when the ${\cal T}$ symmetry is broken by Coulomb interactions which generate magnetically ordered states whose spatially averaged magnetization vanishes in the absence of relativistic spin-orbit coupling.
When materials in this class have a finite Hall conductivity, we will refer to them as anomalous Hall antiferromagnets.  As we will discuss,
the refined definition of zero averaged magnetization in the absence of spin-orbit coupling is relevant, since the crystal symmetries consistent with an anomalous Hall effect
also allow a small but non-zero spatially averaged magnetization when spin-orbit coupling is turned on.
One important characteristic of anomalous Hall antiferromagnets is that the Hall vector can be allowed or excluded by symmetry depending on the orientation of the 
magnetic-order (N\'eel) vector. In contrast, ferromagnets have  an anomalous Hall vector allowed for any orientation of ${\bf M}$.

\smallskip

\noindent{\bf\em Non-relativistic magnetic order and time-reversal symmetry breaking in antiferromagnets.} We first analyze the symmetry conditions for the Hall effect in crystals composed of two sublattices.  They can contain, besides  magnetic atoms, also non-magnetic atoms, and the corresponding elements between the two sublattices are chemically equivalent. Among those, we focus on crystals whose 
nuclear-position (charge) structure has at least one spatial symmetry operation (translation, inversion, rotation, or a combination of these) that
transposes one sublattice onto the other -- a sublattice-transposing symmetry.\cite{Landau1984,Turov1965}

Non-relativistic electron-electron Coulomb interactions can lead to quantum ground states 
that have non-zero magnetic moments on the two sublattices.  The transposing symmetry  then favors states with magnetic moments of precisely equal magnitudes.
 Ferromagnetic order with a strong non-relativistic magnetization corresponds to a parallel alignment of the sublattice moments. 
An anti-parallel order, on the other hand, gives a strictly zero net magnetization in the absence of spin-orbit coupling.\cite{Landau1984,Turov1965} 
In the following section we will see that when the sublattice-transposing symmetry is a crystal translation or inversion, the Hall effect is excluded by symmetry, as commonly assummed in antiferromagnets. However, unexpectedly rich physics, including the possibility of a Hall effect, opens up when the sublattice-transposing symmetry contains a crystal rotation.\cite{Smejkal2020,Feng2020a,Ahn2019,Gonzalez-Hernandez2021,Reichlova2020,Smejkal2021,Shao2021,Smejkal2021a} 

Note, that in the absence of the transposing symmetry,  the sublattice moments  will generically have unequal magnitudes and even an anti-parallel coupling of the sublattice moments 
will generate a net non-relativistic magnetization. In systems of this type, commonly referred to as ferrimagnets, the non-relativistic magnetization  
can be comparable in strength to that of ferromagnets. Ferrimagnets are not addressed further in this review.\cite{Shi2018} 
 
\smallskip
\noindent{\bf\em Spin-orbit coupling and Hall vector.} The Hall vector is an orbital response function.  
When the relativistic spin-orbit coupling terms are included in the Hamiltonian, ${\cal T}$-symmetry breaking in the
spin-space can by macroscopically probed by the Hall response.\cite{Nagaosa2010,Chen2014,Kubler2014,Smejkal2020}
Two-sublattice magnetic crystals with antiparallel sublattice moments are described by the N\'eel vector ${\bf L}={\bf M}_1-{\bf M}_2$, where ${\bf M}_{1(2)}$ is the magnetic moment of the first (second) sublattice. 
As we explain in detail below, a Hall response can occur when the presence of the N\'eel vector and spin-orbit coupling lowers the symmetry of the system sufficiently to allow for a ${\cal T}$-odd axial vector. Before analyzing these symmetry conditions, we first recall how to apply symmetry operations in the presence of spin-orbit coupling, and then we inspect the symmetry transformation rules for the N\'eel vector.\cite{Landau1984,Turov1965}

When we discussed magnetic order in the absence of relativistic spin-orbit coupling above, 
we assumed invariance under mutual rotations 
of charge and spin spaces, i.e. that the spin and charge sectors were uncoupled, and implicitly recognized that 
spin is invariant under translation and inversion.  The sublattice-transposing symmetry operations were then considered only in the crystal charge space.
In the presence of spin-orbit coupling, however, the spin and charge spaces are coupled, and rotation symmetry operations have to be applied jointly in both spaces. 

From the above definition of the  N\'eel vector we see that its symmetry transformation rules include those of a ${\cal T}$-odd axial vector since ${\bf M}_{1(2)}$ are ${\cal T}$-odd axial vectors. 
On top of that, however, the N\'eel vector also flips if the symmetry operation is sublattice-transposing. 

When spin-orbit coupling is included, a sublattice-transposing symmetry may persist  or 
may be broken. If all the sublattice-transposing symmetries are removed when spin-orbit coupling is included,
the N\'eel vector transforms as a ${\cal T}$-odd axial vector,  which means that the Hall vector is then allowed by symmetry. 
On the other hand, when the sublattice-transposing symmetry is retained in the presence of spin-orbit coupling, the N\'eel vector does not transform as an axial vector under this symmetry (recall the extra sign change). In this case, the Hall effect may or may not be allowed. The result depends on the specific symmetries of the magnetic crystal, which we inspect more closely in the following paragraphs.

\smallskip

\noindent{\bf\em Translation or inversion sublattice-transposing symmetry.}
When the two-sublattice collinear antiferromagnetic state is formed by doubling the unit cell of the underlying charge crystal, the antiferromagnet has a {\em t}$_{1/2}{\cal T}$-symmetry, where {\em t}$_{1/2}$ is a half-unit cell translation of the antiferromagnetic lattice. This is illustrated in Fig.~2a. Since {\em t}$_{1/2}$ is a sublattice-transposing symmetry operation, and an axial vector is {\em t}$_{1/2}$-invariant, the N\'eel vector  ${\bf L}$ flips sign under {\em t}$_{1/2}$. ${\cal T}$ also reverses ${\bf L}$, which makes the N\'eel vector invariant under the {\em t}$_{1/2}{\cal T}$-symmetry operation. This confirms that a {\em t}$_{1/2}{\cal T}$-symmetric antiferromagnetic crystal can exist. A ${\cal T}$-odd axial Hall vector, on the other hand, is odd under {\em t}$_{1/2}{\cal T}$ and, therefore, excluded in {\em t}$_{1/2}{\cal T}$-symmetric antiferromagnets.

Next we look at antiferromagnetic crystals with a ${\cal PT}$-symmetry, where ${\cal P}$ is spatial inversion. An illustrative example is shown in Fig.~2b. Since ${\cal P}$ does not flip the
sign of an axial vector, the ${\cal P}$-symmetry operation is sublattice-transposing in the presence of the ${\cal PT}$-symmetry, which generates one sign change of ${\bf L}$. Another sign change is due to ${\cal T}$ and the N\'eel vector is, therefore, invariant under the ${\cal PT}$-symmetry operation and can exist in ${\cal PT}$-symmetric crystals. The Hall vector, on the other hand,  is odd under ${\cal PT}$ and is again excluded. 

\newpage

\onecolumngrid

\begin{figure}[h!]
\vspace{1cm}
\hspace*{-0cm}\epsfig{width=.8\columnwidth,angle=0,file=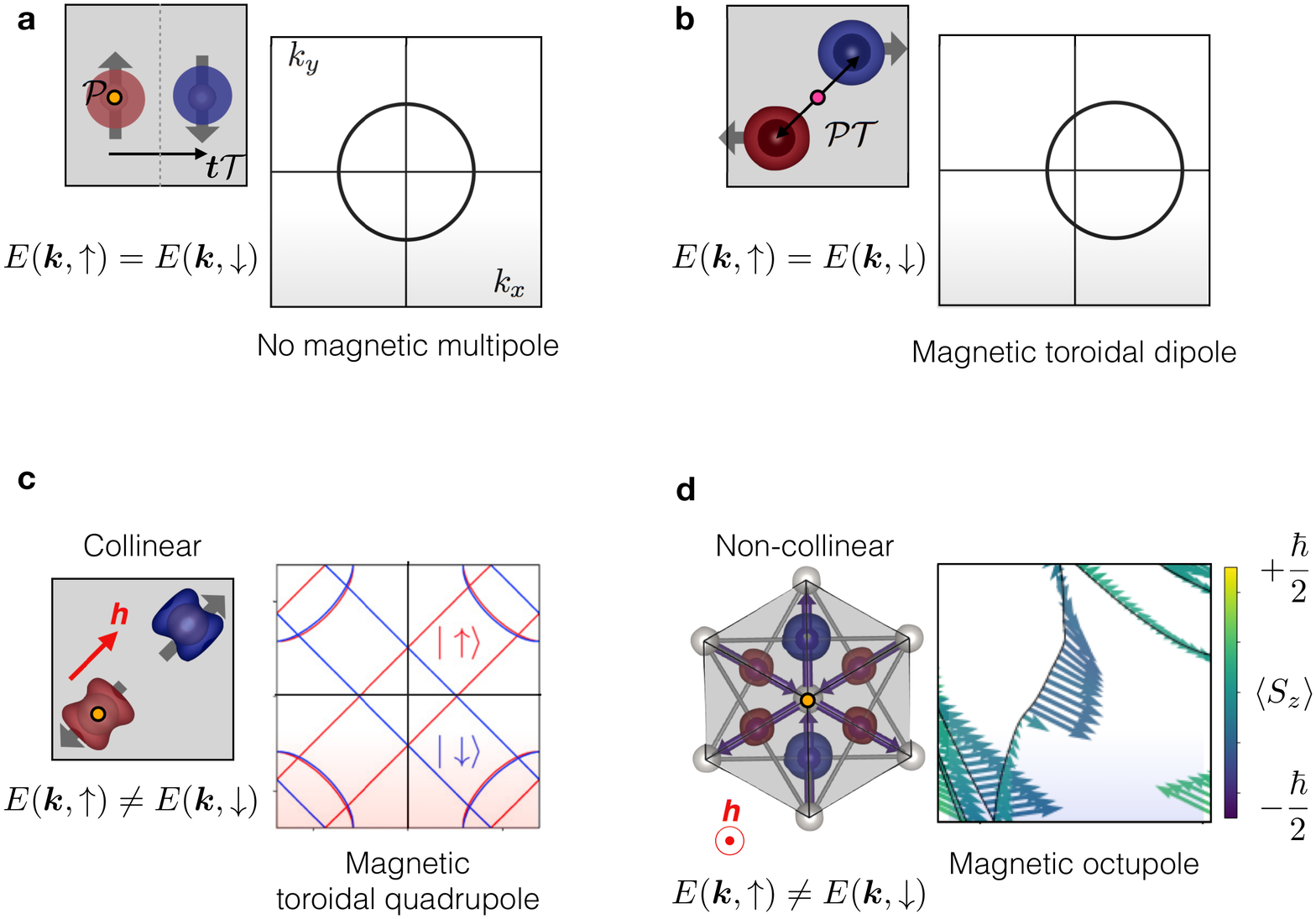}
\caption{{\bf Collinear and non-collinear archetype structures of anomalous Hall antiferromagnets, and magnetic multipoles.} 
{\bf a,} An antiferromagnet with a {\em t}$_{1/2}{\cal T}$ sublattice-transposing symmetry, no magnetic multipole, and no Hall effect. 
If combined with ${\cal P}$-symmetry, the bands are spin-degenerate. {\bf b,} An antiferromagnet with a ${\cal PT}$ sublattice-transposing symmetry, magnetic toroidal dipole, spin-degenerate bands, and no Hall effect. {\bf c,} A compensated collinear magnet with non-relativistic alternating spin-polarization in the momentum space and corresponding magnetic toroidal quadrupole Fermi surface
that generates a Hall effect when spin-orbit coupling is included. {\bf d,} A compensated non-collinear coplanar magnet with a non-relativistic spin-texture in momentum space due to the non-collinear magnetic order, with a magnetic octupole, that generates a Hall effect when spin-orbit coupling is included.
}
\label{fig2}
\end{figure}

\newpage

\twocolumngrid

\smallskip

\noindent{\bf\em Rotation sublattice-transposing symmetries.} Unlike {\em t}$_{1/2}{\cal T}$ and ${\cal PT}$-symmetries, rotation transposing symmetries do not necessarily exclude the Hall vector. Whether or not  the Hall vector is allowed depends on other symmetries in the magnetic space group. As an illustration, we show in Fig.~3a a specific example of a text-book two-sublattice rutile structure which, besides magnetic atoms, contains also non-magnetic elements in the lattice.\cite{Tinkham2003,Smejkal2020} Here the non-relativistic collinear anti-parallel order has a zero net magnetization since the magnetic moments of the two sublattices have strictly equal magnitudes due to either of the two   (screw-axis) rotation transposing-symmetries: {\em t}$_{1/2}{\cal C}_{2x}$ and {\em t}$_{1/2}{\cal C}_{2y}$. Here ${\cal C}_{2x(y)}$ are 2-fold rotation-axis symmetries. 

When including spin-orbit coupling and aligning the N\'eel vector ${\bf L}$ along the $z$-axis (magnetic space group {\em P}$4^\prime_2$/{\em mnm$^\prime$}), the {\em t}$_{1/2}{\cal C}_{2x}$ and {\em t}$_{1/2}{\cal C}_{2y}$ symmetries are retained. This excludes the Hall vector because a vector cannot be simultaneously invariant under two orthogonal rotation symmetry axes. On the other hand, for ${\bf L}$ parallel to the $x$-axis (magnetic space group Pn{\em n$^\prime$m$^\prime$}), the {\em t}$_{1/2}{\cal C}_{2y}$ is retained while the other rotation transposing-symmetry changes to {\em t}$_{1/2}{\cal C}_{2x}{\cal T}$. The ${\cal T}$-odd Hall vector is now allowed along the $y$-axis, i.e., it is orthogonal to ${\bf L}$ (Fig.~3a). 

\newpage

\onecolumngrid

\begin{figure}[h!]
\hspace*{-0cm}\epsfig{width=.8\columnwidth,angle=0,file=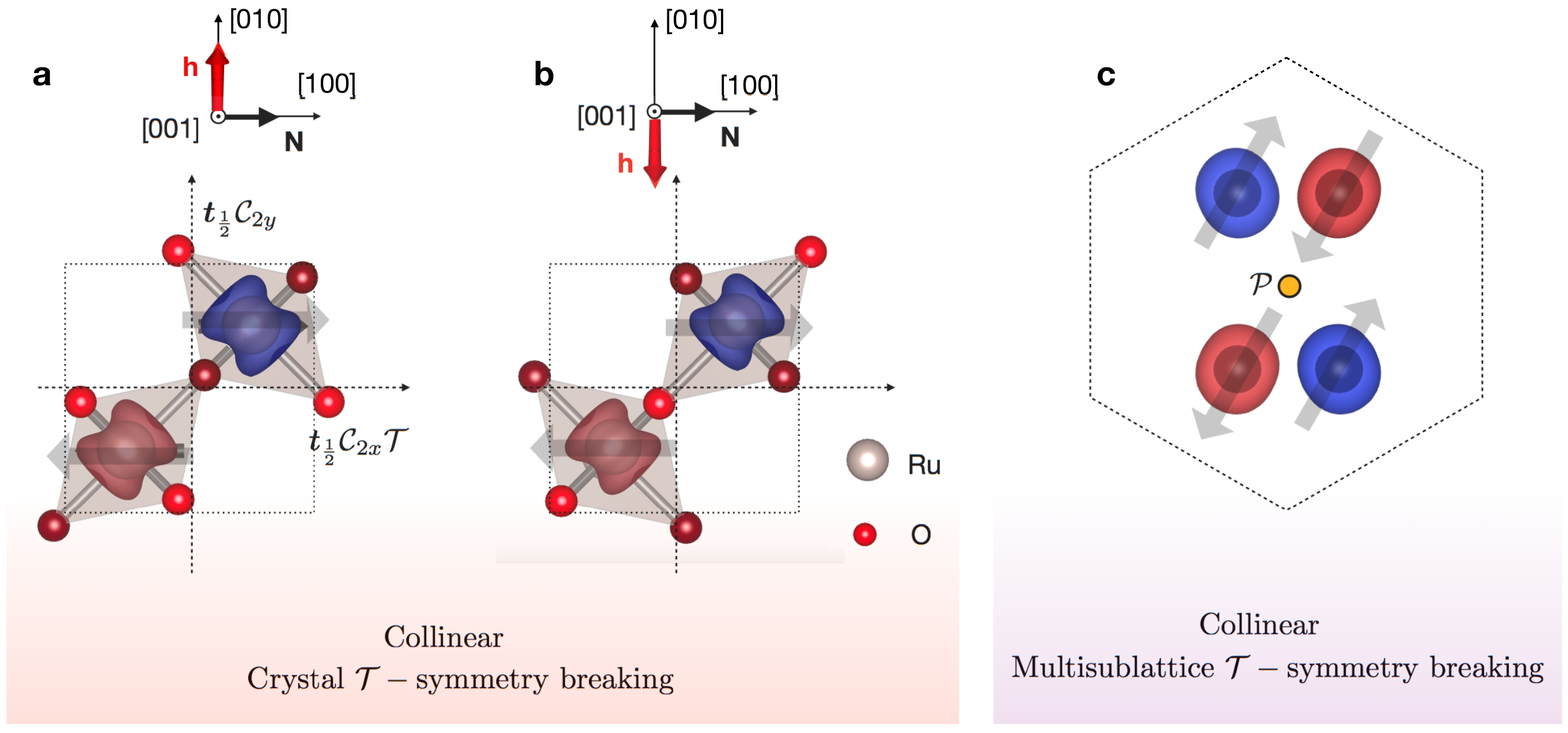}
\caption{{\bf Typical compensated collinear structures with multisublattice  ${\cal T}$-symmetry breaking.}
{\bf a,} A crystal with rotation sublattice-transposing symmetries and ${\cal T}$-symmetry breaking by collinear antiparallel magnetic order and anisotropic crystal environments that 
generate a Hall effect.  {\bf b,} In this mechanism, the sign of the  Hall effect can be flipped by reversing the local crystal anisotropies while keeping the sublattice moment directions
and the N\'eel vector fixed. {\bf c,} Breaking of the ${\cal PT}$ symmetry by a compensated four-sublattice collinear order in which the sublattices transposed one onto the other by ${\cal P}$ have parallel moments.
}
\label{fig3}
\end{figure}

\twocolumngrid

Another scenario is when ${\bf L}$ is aligned with the $xy$-plane diagonal axis (magnetic space group Cm{\em m$^\prime$m$^\prime$}). In this case, the transposing symmetries are removed, 
and the N\'eel vector transforms as a ${\cal T}$-odd axial vector under all remaining symmetries since these are non-transposing. The Hall vector is then allowed. In the specific magnetic symmetry space group of the considered rutile structure for ${\bf L}$ along the in-plane diagonal, the Hall vector ${\bf h}\parallel{\bf L}$ (Fig.~2c). 

We see that in the same magnetic crystal, the Hall vector can be excluded or allowed by symmetry, depending on the N\'eel vector orientation. 
This makes the anomalous Hall antiferromagnets distinct from ferromagnets, in which the Hall vector is always allowed by symmetry for any orientation of the magnetization.

\smallskip

 \noindent{\bf\em Symmetry rules for the anomalous Hall antiferromagnets.} Next we review the magnetic symmetry groups that allow Hall vectors, and  specify the Hall vector orientation determined by the magnetic crystal symmetry. Representative examples of  anomalous Hall antiferromagnets for each of the symmetry groups are given in, e.g.,  Ref.~\onlinecite{Smejkal2020} (see also Tab.~1 below). We  emphasize that the symmetry group of a given magnetic crystal depends also on the orientation of the magnetic moments with respect to the crystal axes. 
 
The symmetry rules for the Hall vector are as follows:

\begin{itemize} 

\item 
We study the bulk Hall conductivity in metallic systems which is a macroscopic spatially averaged quantity, i.e., invariant under translation. We can, therefore, consider only magnetic point groups which are obtained from the magnetic space groups by regarding every translation as an identity, i.e., also regarding the screw rotational  operations are simple rotations, and glide mirror planes as simple mirror symmetries.\cite{Landau1984} Only 31 magnetic point groups allow for a Hall vector.\cite{Kleiner1966,Seemann2015,Smejkal2020}

\item
Since both current and electric field flip sign under ${\cal P}$, the 
linear-response
 conductivity tensor is ${\cal P}$-invariant. We can  therefore limit the discussion to magnetic Laue groups obtained from the magnetic point groups by regarding inversion as 
 an identity and the mirror planes as two-fold rotations around axes orthogonal to the planes. Beside identity, the only remaining symmetry operations in the magnetic Laue groups are then  ${\cal T}$, rotations, and their combinations.  Only 10 magnetic Laue groups allow for a Hall vector.\cite{Kleiner1966,Seemann2015,Smejkal2020} 

\item 
The lowest symmetry Laue group contains just the identity element. In this case, symmetry does not prescribe any specific orientation of the Hall vector. The orientation is then determined purely by the microscopic properties of the material. 

\item
A magnetic Laue group with only one (beside identity) symmetry element 2$^\prime$, which is a combination of  a two-fold rotation and ${\cal T}$, restricts the Hall vector to the plane orthogonal to the rotation axis. 

\item
In all remaining eight magnetic Laue groups labeled as, 2, 3, 4, 6, 2$^\prime$2$^\prime$2, 42$^\prime$2$^\prime$, 32$^\prime$, and 62$^\prime$2$^\prime$, the Hall vector is along the rotation axis not combined with  ${\cal T}$ (unprimed).
\end{itemize}

\smallskip

\noindent{\bf\em Archetype two-sublattice and multiple-subllattice structures.} 
Antiferromagnets  containing only two magnetic atoms with antiparallel moments in the unit cell  and no other elements in the crystal have at least ${\cal PT}$-symmetry, and additionally can also have {\em t}$_{1/2}{\cal T}$-symmetry. A Hall vector  is excluded by either of these symmetries, and the ${\cal PT}$-symmetry also guaranties that the 
energy bands are Kramers spin-degenerate over the entire Brillouin zone. The ${\cal PT}$-symmetric antiferromagnets can, nevertheless, have  a magnetic (polar) toroidal dipole (Fig.~2b),\cite{Watanabe2018a,Hayami2018,Thole2020} and can allow for  ${\cal T}$-odd second-order magneto-transport phenomena,\cite{Watanabe2018a,Godinho2018} and for a field-like N\'eel spin-orbit torque and associated electrical switching of the antiferromagnetic order vector.\cite{Zelezny2014,Watanabe2018a}  CuMnAs  or Mn$_2$Au are prominent material examples  with metallic conduction and high N\'eel temperature falling into this class.\cite{Zelezny2014,Wadley2016,Bodnar2018,Watanabe2018a} 

In anomalous Hall antiferromagnets with two collinear antiparallel sublattices, the magnetic atoms have to be complemented by additional (non-magnetic) atoms in the unit cell.\cite{Smejkal2020} The rutile structure in Fig.~3a is an example showing explicitly how the non-magnetic atoms contribute to breaking the {\em t}$_{1/2}{\cal T}$ and ${\cal PT}$-symmetries. 
Alternatively, as shown in Fig.~2c, the symmetry breaking can be also seen by focusing only on the magnetic atoms if the real-space anisotropic magnetization densities are plotted on top of the positions of the magnetic atoms and their moment orientations. The momentum-space counterpart is an anisotropic altermagnetic\cite{Smejkal2021a} spin-polarization of the non-relativistic energy bands, with zero net magnetization.\cite{Smejkal2020,Ahn2019,Hayami2019,Feng2020a,Gonzalez-Hernandez2021,Smejkal2021a} The corresponding Fermi surfaces can have the characteristic symmetry of a magnetic toroidal quadrupole (Fig.~2c).\cite{Hayami2018,Hayami2020} (See next section for more details on the electronic structure.)


A remarkable feature of the Hall effect in this archetype structure is that it flips sign not only when reversing ${\bf L}$ but also when the symmetry-breaking arrangement of non-magnetic atoms reverses between the two magnetic sublattices while keeping the same sign of ${\bf L}$ (see Figs.~3a,b). Alternatively, the Hall effect can be turned on and off by changing the relative angle of the non-magnetic structures surrounding the  magnetic atoms in the two sublattices, i.e., by turning off and on the {\em t}$_{1/2}{\cal T}$-symmetry. Because of this sensitivity to the crystal environment and crystal fields,\cite{Smejkal2021a} the mechanism has been referred to as the crystal Hall effect.\cite{Smejkal2020,Samanta2020} RuO$_2$ is a representative room-temperature metallic member of this family of anomalous Hall antiferromagnets.\cite{Berlijn2017a,Zhu2018,Smejkal2020,Ahn2019,Hayami2019,Feng2020a} 

Next we include in our discussion typical magnetic structures with multiple sublattices. A three-sublattice structure (Fig.~2d), with a subblatice-transposing symmetry in the crystal's charge space of a 3-fold rotation axis, can host magnetic order with equal magnitudes of the sublattice moments in the absence of relativistic spin-orbit coupling.  
Frustrated anti-parallel magnetic coupling  on the triangular lattice can then result in a compensated co-planar non-collinear magnetic order with a 120$^\circ$ tilt-angle between the sublattice moments. Because of the non-collinear order, no additional (non-magnetic) atoms are needed in the lattice in this case to allow for breaking the {\em t}$_{1/2}{\cal T}$ and ${\cal PT}$-symmetries. Depending on the other symmetries of the magnetic crystal for a given orientation of the magnetic moments with respect to the lattice, the Hall vector may be allowed or excluded, in analogy to the above discussion of the two-sublatice magnetic structures. The Hall effect was identified in several members of the family of compensated non-collinear Mn$_3$X (X=Ir,Sn,Ge,Pt) magnets. Instead of a magnetic dipole, these
materials are characterized by a macroscopic magnetic octupole and a non-collinear spin-texture in the momentum space, induced by the real-space non-collinear order even in the absence of the relativistic spin-orbit coupling (Fig.~2d).\cite{Chen2014,Kubler2014,Nakatsuji2015,Kiyohara2015,Nayak2016,Suzuki2017,Higo2018b,Suzuki2019a,Liu2018b,Tsai2020,Taylor2020,Nomoto2020} 

Four magnetic moments in a unit cell can be arranged in a compensated antiparallel order without frustration in their magnetic couplings.  
If the sublattices transposed one onto the other by space inversion have parallel moments (Fig.~3c),  the  ${\cal PT}$-symmetry is broken, opening a possibility for the Hall effect. This has been demonstrated in a compensated four-sublattice checkerboard magnetic phase of Mn$_5$Si$_3$.\cite{Reichlova2020}  Like the two-sublattice Hall-effect archetype, its non-relativistic electronic structure has the altermagnetic\cite{Smejkal2021a} collinear spin-polarization in momentum space,  generating  zero net magnetization.\cite{Reichlova2020} 

Finally, we mention a possibility of a compensated order with non-collinear non-coplanar moments in a four-sublattice structure. Remarkably, this structure can in principle allow for a Hall effect without relativistic spin-orbit coupling, \cite{Taguchi2001,Shindou2001,Machida2010,Ueland2012,Surgers2014,Suzuki2017} in which case the Hall effect can be invariant under independent rotations of the spin-space. However, because of the non-coplanarity,  the ${\cal T}$ operation is not equivalent to a 180$^\circ$ rotation of the spin space. The absence of a symmetry combining the operations of ${\cal T}$ and 180$^\circ$ rotation of the spin space illustrates the possibility of  the ${\cal T}$-odd Hall effect in a non-coplanar compensated magnet in the absence of the relativistic spin-orbit coupling.\cite{Suzuki2017} As in all systems, the non-coplanar magnetic crystal has to also break all other symmetries consistent with the presence of an axial vector.

\bigskip

\smallskip

\noindent{\bf\em Weak ferromagnetism.}  Above we discussed the Hall effect in the framework of magnetic crystal symmetries while in the next section we will elaborate on its microscopic origin due to the relativistic Berry-phase. Before turning to the microscopics, we make a comment here on notable equilibrium relativistic  effects which may also occur in the anomalous Hall antiferromagnets. 

One example is the magnetocrystalline anisotropy, which can be present in any magnetically ordered system, and  which tends to align the N\'eel vector along specific crystal direction(s). The  thermodynamic potential of the magnet then includes, besides the strong isotropic magnetic coupling terms,  also weak relativistic  terms  which are anisotropic in ${\bf L}$.\cite{Landau1984} 

Another example of equilibrium relativistic effects is a non-zero spatially averaged magnetization, ${\bf M}={\bf M}_1+{\bf M}_2$.  Microscopically, it can originate from a spin-orbit coupling induced Dzyaloshinskii-Moriya interaction. It adds to the thermodynamic potential additional weak relativistic terms which couple ${\bf L}$ and ${\bf M}$.\cite{Landau1984} In the above discussed rutile crystal with ${\bf L}$ parallel to the $y$-axis, i.e. when the transposing symmetry  {\em t}$_{1/2}{\cal C}_{2x}$ is retained, the magnitudes of  ${\bf M}_1$ and ${\bf M}_2$ remain the same even in the presence of spin-orbit coupling. A week relativistic ${\bf M}$ is allowed by symmetry in a direction orthogonal to  ${\bf L}$ and corresponds to a small canting of the antiparallel moments towards the $x$-axis. In 
the literature, this relativistic scenario is commonly referred to as weak ferromagnetism. For ${\bf L}$ aligned with the $xy$-plane diagonal, when all sublattice-transposing symmetries are removed by spin-orbit coupling, the weak relativistic ${\bf M}$ is due to slightly unequal magnitudes of the sublattice magnetizations and, correspondingly, ${\bf M}\parallel {\bf L}$. The system can then be regarded as a weak ferrimagnet.

Since ${\bf M}$ is a ${\cal T}$-odd axial vector, weak ferromagnetism (ferrimagnetism) is allowed or excluded following the same symmetry conditions as the Hall vector ${\bf h}({\bf L})$.  We emphasize, however, that ${\bf h}({\bf L})$ is not driven by the relativistic  weak ferromagnetism (ferrimagnetism). As shown above, the  Hall effect originates from ${\cal T}$-symmetry breaking due to the  ${\bf L}$-order, which is induced by non-relativistic Coulomb interactions of a similarly large strength as the non-relativistic ${\bf M}$-order in ferromagnets. In both case, the relativistic  spin-orbit interaction  then couples  the ${\cal T}$-symmetry breaking to the charge Hall transport. 

Because of the presence of the relativistic weak ferromagnetic moment in  anomalous Hall antiferromagnets, the precise application of Onsager relations implies that ${\bf h}({\bf L,M})={\bf -h}({\bf -L,-M})$. However, the magnitude of  the Hall effect can be driven primarily by  the strong non-relativistic ${\bf L}$-order, while weak relativistic ${\bf M}$ causes only a small correction.\cite{Nakatsuji2015,Nayak2016,Smejkal2020,Feng2020a} 



\subsection*{Hall effect and Berry curvature}
We now proceed to the microscopic description of the Hall effect. We limit our discussion to the intrinsic disorder-independent contribution. This focus is
 motivated by previous studies of metallic ferromagnets in which the intrinsic anomalous Hall conductivity is often the largest contribution within  a typical range of longitudinal conductivities of $\sim 10^4-10^6$~$\Omega^{-1}{\rm cm}^{-1}$.\cite{Nagaosa2010}
An extensive discussion of disorder effects in the anomalous Hall effect can be found in, e.g., Ref.~\onlinecite{Nagaosa2010}. 
We start this section by  making a few qualitative remarks on the connection between the intrinsic contribution to the Hall effect and the Berry phase.

\smallskip

\noindent{\bf\em Berry phase.} The intrinsic contribution to the antisymmetric Hall conductivity is given by the Kubo formula,\cite{Marder2010,Tong2016}
\begin{eqnarray}
&&\sigma^{\rm a}_{xy}=\frac{e^2}{\hbar}\sum_{n'\neq n}\int_{\rm BZ} \frac{d^3k}{(2\pi)^3} f[\varepsilon_n({\bf k})] \nonumber \\
&\times&\frac{2{\rm Im}[\langle u_{n{\bf k}}|\partial_{k_x}\hat H({\bf k})|u_{n'{\bf k}}\rangle \langle u_{n'{\bf k}}|\partial_{k_y}\hat H({\bf k})|u_{n{\bf k}}\rangle]}{[\varepsilon_n({\bf k})-\varepsilon_{n'}({\bf k})]^2}\ .
\label{Kubo}
\end{eqnarray}
Here $f[\varepsilon_n({\bf k})]$ is the Fermi-Dirac distribution function, $\varepsilon_n({\bf k})$ is the energy of the equilibrium Bloch state in band $n$ with crystal quasi-momentum {\bf k}, $u_{n{\bf k}}({\bf r})$ is the periodic part of the Bloch wavefunction diagonalizing the crystal Hamiltonian $\hat H({\bf k})$,
and  $\nabla_{\bf k}\hat H({\bf k})/\hbar$ is the velocity operator.

The Kubo formula (\ref{Kubo}) can be rewritten by noting that 
$\nabla_{\bf k}\langle u_{n{\bf k}}|\hat H({\bf k})|u_{n'{\bf k}}\rangle=0$ and $\nabla_{\bf k}\langle u_{n{\bf k}}|u_{n'{\bf k}}\rangle=0$ 
from which it follows that\cite{Berry1984a,Nagaosa2010,Xiao2010b}
\begin{equation}
\frac{\langle u_{n{\bf k}}|\nabla_{\bf k}\hat H({\bf k})|u_{n'{\bf k}}\rangle}{\varepsilon_n({\bf k})-\varepsilon_{n'}({\bf k})}=\langle\nabla_{\bf k}u_{n{\bf k}}|u_{n'{\bf k}}\rangle
\end{equation}
and 
\begin{equation}
\sigma^{\rm a}_{xy}=-\frac{e^2}{\hbar}\sum_{n}\int_{\rm BZ} \frac{d^3k}{(2\pi)^3} f[\varepsilon_n({\bf k})] {\mathcal{B}}_n^z({\bf k}) ,
\label{Berry}
\end{equation}
where
\begin{eqnarray}
\bm{\mathcal{B}}_n({\bf k})&=&\nabla_{\bf k}\times\bm{\mathcal{A}}_n({\bf k})\nonumber \\
\bm{\mathcal{A}}_n({\bf k})&=&i\langle u_{n{\bf k}}|\nabla_{\bf k}u_{n{\bf k}}\rangle .
\label{Berry-con-cur}
\end{eqnarray}

Eqs.~(\ref{Kubo})-(\ref{Berry-con-cur}) show a direct connection between the Hall effect and a general concept of the geometrical Berry phase.\cite{Thouless1982,Berry1984a,Simon1983,Kohmoto1985,Nagaosa2010,Xiao2010b,Hasan2010,Tong2016} Berry's concept is based on considering a non-degenerate energy eigenstate evolving adiabatically when parameters of the system's Hamiltonian change slowly in time.  
Apart from a  familiar dynamical phase proportional to the time integral of energy, the wavefunction can acquire an additional phase factor. When completing a closed loop in the parameter space, which in Eqs.~(\ref{Kubo})-(\ref{Berry-con-cur}) is represented by the crystal momentum space, this additional factor is called the Berry phase.\cite{Berry1984a} Since it is given by a path integral of the Berry connection $\bm{\mathcal{A}}_n({\bf k})$, i.e. is independent of the rate of the adiabatic parameter change, the Berry phase is geometrical. 

Using Stoke's theorem, the Berry phase can be expressed as an integral of the Berry curvature $\bm{\mathcal{B}}_n({\bf k})$ over an area enclosed by the loop.
The Berry phase concept applies in general to any parameter space.  A prominent example is the Aharonov-Bohm phase in the real-space.\cite{Berry1984a}  $\bm{\mathcal{A}}_n({\bf k})$ (and $\bm{\mathcal{B}}_n({\bf k})$) in Eq.~(\ref{Berry-con-cur})  can be viewed as a {\bf k}-space analogue of  the electrodynamic vector potential (and magnetic field) generating the Aharonov-Bohm phase. 

\smallskip

\noindent{\bf\em Berry-phase quantum and anomalous Hall effects.}  We first recap some of the key observations made in earlier studies of the quantum and anomalous Hall effects induced by the magnetic dipole.


When applying the $\mathcal{T}$-symmetry operation on the Berry curvature in Eq.~(\ref{Berry-con-cur}) we get, $\mathcal{T} \bm{\mathcal{B}}_n({\bf k})=-\bm{\mathcal{B}}_n({\bf -k})$ since ${\cal T}$ is antiunitary, i.e., includes complex conjugation. This implies that in $\cal{T}$-invariant systems, $\bm{\mathcal{B}}_n({\bf k})=-\bm{\mathcal{B}}_n({\bf -k})$, and the integral of $\bm{\mathcal{B}}_n({\bf k})$ vanishes, consistent with a vanishing Hall effect. 

In analogy to Gaussian curvature, the integral over any closed 2D surface of the Berry curvature is an integer multiple of $2\pi$ where the integer $C$ is called the Chern topological invariant.\cite{Simon1983,Hasan2010,Tong2016} A 2D Brillouin zone is an example of the closed 2D surface. It is topologically equivalent to a torus since states with crystal quasi-momenta separated by a reciprocal lattice vector are identical. 

A quantum Hall system with an integer Landau level filling factor\cite{Klitzing1980} corresponds to an insulating 2D crystal with a unit cell area enclosing a magnetic flux quantum.\cite{Thouless1982,Simon1983,Kohmoto1985,Hasan2010,Marder2010,Tong2016} The integral of the Berry curvature over the 2D Brillouin zone then gives a quantized Hall conductance $\sigma^{2D}_{xy}=C\frac{e^2}{h}$. 

A direct extension to 3D insulators implies that  the Hall conductivity is semi-quantized, given by  $\sigma_{xy}=G_z\frac{e^2}{h}$, where $G_z$ is the $z$-component of a reciprocal lattice vector.\cite{Halperin1987,Tang2019a} 

In metallic systems, the Berry curvature Hall conductivity has a non-quantized part which in ferromagnets is referred to as the intrinsic anomalous Hall effect.\cite{Jungwirth2002,Yao2004,Haldane2004,Nagaosa2010,Xiao2010b} 
A strong contribution to the Berry curvature occurs near band (anti)crossings. In  ferromagnets these  are typically accidental, but their presence and position in the Brillouin zone can be also imposed by crystal symmetry.\cite{Falicov1968,Cracknell1970,Gosalbez-Martinez2015,Kim2018d,Liu2018c,Sakai2018} A prominent example of band crossings are Weyl points which always come in pairs and together act as momentum-space Berry curvature counterparts of real-space magnetic dipoles.\cite{Fang2003,Vafek2013,Witten2016,Felser2017,Kuroda2017,Yang2017c,Noky2019a,Smejkal2020}  

The Hall conductivity tends to be dominated by the Berry curvature from (anti)crossings near the Fermi level. This can be seen when approximating the sum in  Eq.~(\ref{Kubo})  by including only the two nearby bands close to a band (anti)crossing. The Berry curvature  is equal in magnitude but has opposite signs for each band, and a contribution will occur only for the parts of the Brillouin zone where one is occupied and the other one is not. Alternatively, this can also be deduced by rewriting in  Eq.~(\ref{Kubo}) the factor $2 f[\varepsilon_n({\bf k})]$ as $(f[\varepsilon_n({\bf k})]-f[\varepsilon_{n'}({\bf k})])$, which is mathematically equivalent.\cite{Marder2010,Nagaosa2010}



As an illustrative example of the Hall effect in metallic ferromagnets we show in Fig.~4a-c energy bands and the Berry curvature of a model two-band 2D system with a Hamiltonian $H=H_{\rm k}+H_{\rm R}+H_{\rm Z}$, where the kinetic energy term, $H_{\rm k}({\bf k})=2t\left( \cos {k_{x}}+ \cos {k_{y}} \right)\boldsymbol{1}$, the relativistic Rashba spin-orbit coupling term,  $H_{\rm R}({\bf k})=\lambda\left( \sin {k_{x}}\sigma_{y}- \sin {k_{y}\sigma_{x}} \right)$, the non-relativistic momentum-independent spin-coupling term due to the ferromagnetic order, $H_{\rm Z}=\Delta \sigma_{z}$, and $\boldsymbol\upsigma$ is the vector of Pauli spin matrices.
Around the $\boldsymbol\Gamma$-point we can expand the  Hamiltonian as,
\begin{equation}
H(\boldsymbol\Gamma,{\bf k})=4t +t  k^{2} + \lambda \left( k_{x}\sigma_{y} - k_{y} \sigma_{x}  \right) + \Delta \sigma_{z},
\label{Rashba-Gamma}
\end{equation}
and obtain the Berry curvature for the two bands as,\cite{Culcer2003,Nunner2007,Dugaev2008,Nagaosa2010,Xiao2010b}
\begin{equation}
\mathcal{B}(k)_{\pm}=\mp\frac{\lambda^2\Delta}{2(\lambda^2k^2+\Delta^2)^{3/2}}. 
\end{equation}
In Figs~4b,c we plot the Berry curvature for the lower band.  Since the ferromagnetic coupling is typically much stronger than the spin-orbit coupling, the model gives  a weak and nearly constant $\mathcal{B}(k)$ (Fig.~4b), and a correspondingly weak  Hall conductance. On the other hand, the band (anti)crossing scenario around the $\boldsymbol\Gamma$-point is illustrated in this model in the opposite regime of the ferromagnetic coupling much weaker than the spin-orbit coupling. It shows how a large Hall conductance can arise in ferromagnets from a
Berry curvature strongly peaked at the (anti)crossing (Fig.~4c). 

In the complex band structure of real metallic ferromagnets, many of these, typically accidental, Berry curvature hotspots, contribute
 with one or the other sign,  and add to generate the net intrinsic contribution to the Hall effect. This means that the sign of the Hall effect can vary 
 depending on the detailed microscopic parameters of the system and the magnitude of the Hall effect can accidentally be 
 small for parameters around the sign-change. We point out, however, that the  Hall vector in ferromagnets is not prohibited by 
 symmetry for any magnetization direction.  

\newpage

\onecolumngrid

\begin{figure}[h!]
\vspace{1.5cm}
\hspace*{-0cm}\epsfig{width=.82\columnwidth,angle=0,file=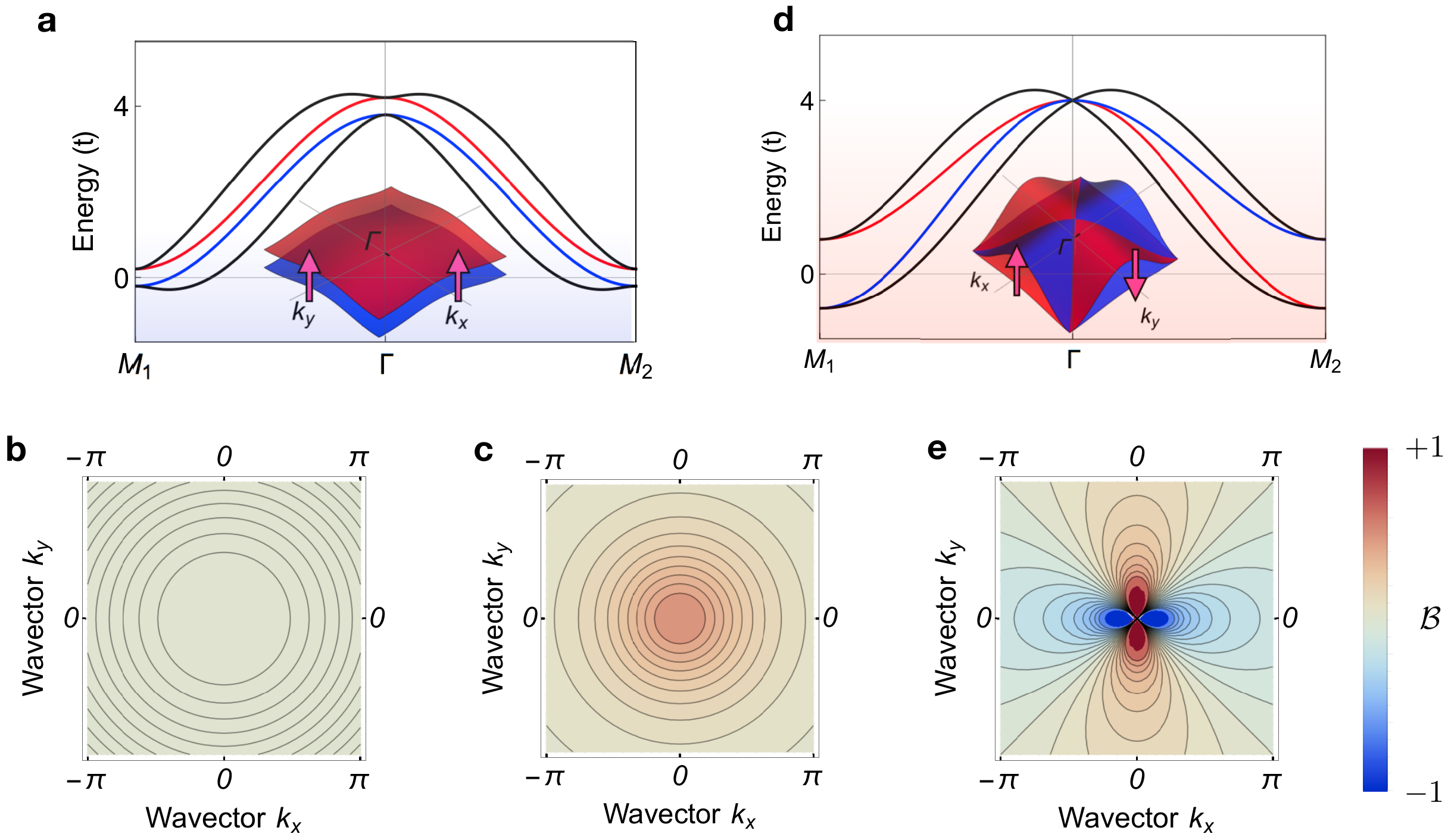}
\caption{{\bf Model Berry curvatures.}
{\bf a}, Model spin-up and spin-down bands (red and blue) split by the non-relativistic ferromagnetic coupling, and band splitting (black) when a strong Rashba spin-orbit interaction is included to generate an (anti)crossing at the $\boldsymbol\Gamma$-point. {\bf b}, A nearly constant weak Berry curvature of the  lower band for weak Rashba spin-orbit coupling. {\bf c}, Berry curvature strongly peaked at the (anti)crossing for strong Rashba spin-orbit coupling. {\bf d}, Model band structure with opposite  non-relativistic spin-splittings in valleys around ${\bf M}_{1}$ and ${\bf M}_{2}$ points. When the Rashba spin-orbit interaction is included (black), the bands still cross at the $\boldsymbol\Gamma$-point which is imposed by the alternating nature of the spin-oplarization in the momentum space. {\bf e}, Corresponding anisotropic Berry curvature around the $\boldsymbol\Gamma$-point band-crossing.
}
\label{fig4}
\end{figure}

\twocolumngrid


\noindent{\bf\em Berry-phase in anomalous Hall antiferromagnets.}
We can now compare the Berry phase phenomenology of ferromagnets and anomalous Hall antiferromagnets.
We start from the collinear magnets  whose non-relativistic bands have the altermagnetic\cite{Smejkal2021a} splitting of spin-up and spin-down bands.\cite{Smejkal2020,Feng2020a,,Gonzalez-Hernandez2021,Reichlova2020,Smejkal2021,Hayami2018,Hayami2019,Ahn2019,Naka2019,Yuan2020,Smejkal2021a} 
The alternating spin-polarization implies that, apart from accidental Berry curvature hotspots analogous to the typical ones in ferromagnets, 
there are band (anti)crossings at the transitions in the Brillouin zone from one sign of spin-splitting to its opposite. These (anti)crossings are thus not accidental, 
but imposed by the symmetry of the altermagnetic spin-polarization in momentum space.\cite{Smejkal2021a}

In Fig.~4d we show an illustrative toy-model example.\cite{Reichlova2020,Smejkal2021}  
The model is obtained by taking the same kinetic and relativistic Rashba terms as used above for a model ferromagnet, 
and replacing the non-relativistic ferromagnetic coupling with an alternating non-relativistic spin-momentum coupling term, $H_{\rm AFZ}=2\Delta \left( \cos {k_{x}}- \cos {k_{y}} \right)\sigma_z$. 
Here the N\'eel vector is set along the $z$-axis. 
Around the $\boldsymbol\Gamma$-point, the Hamiltonian  is given by,
\begin{eqnarray}
H(\boldsymbol\Gamma,{\bf k})&=&4t + t\left(k_{x}^{2}+k_{y}^{2} \right) + \Delta  \left(k_{x}^{2}-k_{y}^{2}\right)\sigma_{z} \nonumber \\
&+&\lambda\left( k_{x}\sigma_{y} - k_{y} \sigma_{x} \right). 
\end{eqnarray}
The Berry curvature near the  $\boldsymbol\Gamma$-point band-crossing is then given by,
\begin{equation}
\mathcal{B}(k)_{\pm}=\mp\frac{\lambda^{2}\Delta\left(k_{x}^{2}-k_{y}^{2}\right)}{\sqrt{\lambda^{2}\left(k_{x}^{2}+k_{y}^{2}\right)+\Delta^{2}\left(k_{x}^{2}-k_{y}^{2}\right)^{2}}}. 
\end{equation}

In Fig.~4e we plot $\mathcal{B}(k)$ for the lower band. It is qualitatively distinct form the Berry curvature of the model ferromagnet shown in Fig.~4c. The Berry curvature around the band-crossing is anisotropic, reflecting  the toroidal quadrupole character of the non-relativistic altermagnetic band structure. Unlike the more isotropic Berry curvature near the ferromagnetic hotspot, Fig.~4e illustrates that the integral of the Berry curvature around the hotspot can vanish. This corresponds to the N\'eel vector orientation prohibiting the Hall vector by symmetry.  As mentioned earlier in this review, the Hall effect can be turned on when the symmetry of the magnetic crystal is lowered by reorienting the N\'eel vector. 

The Berry curvature hotspots associated with the (anti)crossings for the N\'eel vector orientations that allow a Hall vector are seen in relativistic density functional theory (DFT) calculations in Fig.~5.
The results for the rutile magnet RuO$_2$ with two antiparallel sublattices, shown in Figs.~5a-c, again reflect the magnetic toroidal quadrupole symmetry of the non-relativistic altermagnetic band structure  (Figs.~2c and 4d,e). We point out that the spin splitting in RuO$_2$ is on the $\sim$~eV scale, i.e., comparable to spin splittings in typical ferromagnets. 

Figs.~6a-c show results in non-collinear  Mn$_3$Sn, Mn$_3$Ge and Mn$_3$Pt.\cite{Xu2020b,Chen2021a,Liu2018b} 
These figures illustrate the presence of Weyl points near the Fermi level, as well as Berry curvature hotspots at the Fermi surface at accidental band-(anti)crossings. 

In Fig.~6d we also illustrate calculations of a quantized Hall effect for the Fermi level in the band-gap of a 2D Chern insulator realized in a compensated non-coplanar magnetic structure.\cite{Feng2020b}

The DFT calculations  in metallic materials verified that the Hall effect remains virtually unchanged when forcing the weak relativistic magnetization to zero (see Fig.~5d).\cite{Chen2014,Smejkal2020,Feng2020a} This highlights on the microscopic level that the Hall effect is not a consequence of the weak relativistic magnetization. 

Finally, we point out that the Berry curvature mechanism gives quantitatively consistent predictions of the magnitude of the Hall effect 
measured in the metallic compensated magnetic crystals with two and four collinear sublattices, and three non-collinear sublattices.\cite{Smejkal2020,Feng2020a,Chen2014,Kubler2014,Nakatsuji2015,Kiyohara2015,Nayak2016,Zhang2016d,Liu2018b,Suzuki2017,Reichlova2020} 

\newpage

\onecolumngrid

\begin{figure}[h!]
\hspace*{-0cm}\epsfig{width=.8\columnwidth,angle=0,file=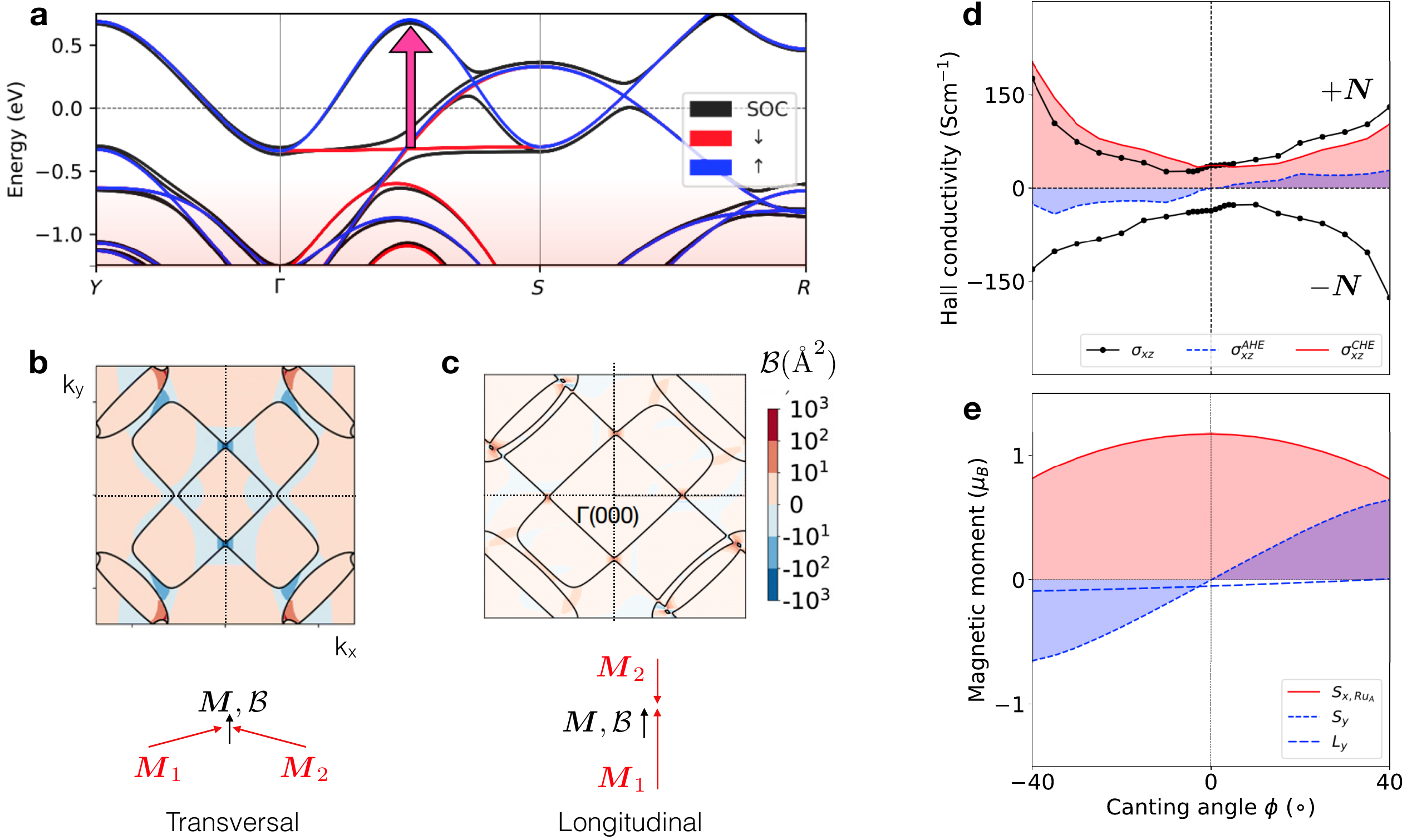}
\caption{{\bf Band structure, Berry curvature and Hall effect calculations in a metallic compensated collinear magnet}.
{\bf a}, Non-relativistic (red and blue) and relativistic (black) bands in RuO$_2$ with an alternating  band splitting on a $\sim$eV scale (highlighted by the arrow). {\bf b}, Berry curvature with hotspots at band (anti)crossings imposed by the alternating spin-polarization in the momentum space. For the N\'eel vector along the [100]-axis, the integrated Berry curvature is along the canting-induced weak relativistic magnetization, i.e., orthogonal  to the N\'eel vector. {\bf c}, Same as b for  the N\'eel vector along the $[1\bar{1}0]$-axis and the integrated Berry curvature is along the weak relativistic magnetization parallel  to the N\'eel vector. {\bf d,e}, Hall conductivity and weak magnetization as a function of the canting angle. Calculations for opposite N\'eel vector and weak magnetization allow to disentangle the small magnetization-induced contribution to the Hall effect from the dominant Hall effect contribution due to the N\'eel vector. Adapted from Refs.~\onlinecite{Smejkal2020}.
}
\label{fig5}
\end{figure}

\twocolumngrid

\onecolumngrid

\begin{figure}[h!]
\hspace*{-0cm}\epsfig{width=.8\columnwidth,angle=0,file=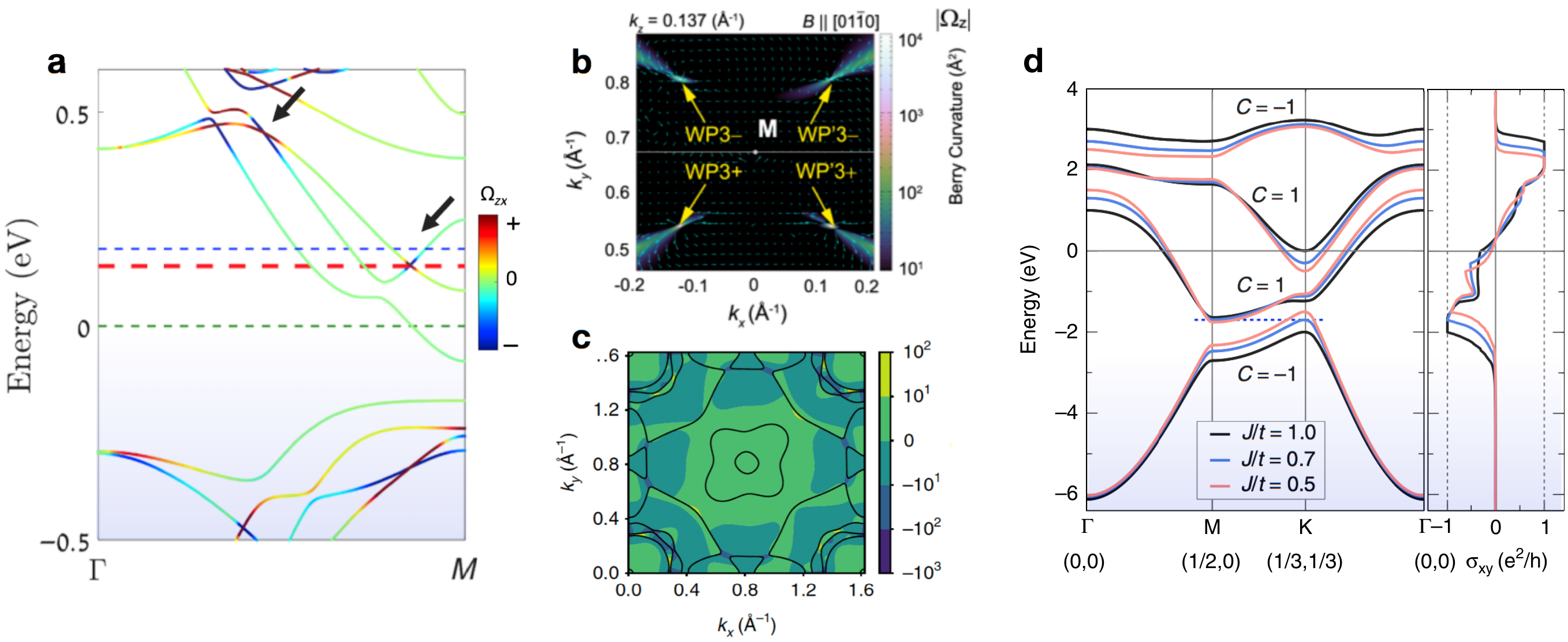}
\caption{{\bf Band structure, Berry curvature and Hall effect calculations in metallic compensated non-collinear magnets and a in compensated non-coplanar magnetic Chern insulator}.
{\bf a}, Band structure of Mn$_3$Sn with the Berry curvature indicated by color. Weyl points are highlighted by arrows.  {\bf b},  Weyl point Berry curvature hotspots in Mn$_3$Ge. {\bf c}, Berry curvature hotspots at the Fermi surface (anti)crossings in Mn$_3$Pt. {\bf d},  Quantized Hall conductance in a model 2D compensated non-coplanar magnet for the Fermi level in a band gap of the Chern insulator. Adapted from Refs.~\onlinecite{Xu2020b,Chen2021a,Liu2018b,Feng2020b}.
}
\label{fig6}
\end{figure}

\twocolumngrid

\clearpage

\onecolumngrid

\subsection*{Summary of experimentally identified materials}



\begin{table}[h!]
\begin{tabular}{clllllll}

\multicolumn{1}{l}{Order}           & System                                               & Synthesis               & $\rho_{H} (\mu\Omega {\rm cm})$ & $\sigma_{AH}$ (S/cm) & T (K) (H (T))        & SG (MSG)          & Note                                                                   \\ \hline \hline
                                            &      Mn$_{3}$Sn   \cite{Nakatsuji2015}         & SC    & 4               & 20               & 300 (0)          & \multicolumn{2}{l}{$P6_{3}/mmc$ ($Cm'cm'$ )}                                                                       \\
                                              &    Mn$_{3}$Ge \cite{Kiyohara2015,Nayak2016}  & SC      & 0.1             & 400              & 5 (0.1)           & $P6_{3}/mmc$ ($C2'/m'$)                   & Mn$_{3}$Ge \cite{Kiyohara2015}                             \\
                                             &     Mn$_{3}$Pt \cite{Liu2018b}                & 20 nm                &                 & 98               & 20 (0)          & $Pm\overline{3}m$ ($R\overline{3}m'$)                                                                           \\
                                              &    Mn$_{3}$Sn \cite{Higo2018b}                                               & 40 nm                & 1.5             & 17               & 300 (0)            & $P6_{3}/mmc$                           &                                                                        \\
 \multirow{-5}{*}{NCAF} & Mn$_{3}$Sn*   \cite{Taylor2020}      & 30 nm epi.           & 0.05            & 21               & 300 (0)          & $P6_{3}/mmc$                            &                                                                        \\ \hline
                                                 & RuO$_{2}$ \cite{Feng2020a}                                                & 27 nm              & 0.05   & 330    & 10 K (30) & $P4_{2}/mnm$ ($Pnn'm'$ ) &                                                             \\
                                                 & Mn$_{5}$Si$_{3}$ \cite{Reichlova2020}         & 12 nm   epi.    & 0.3    & 5     & 110 (0) & \multicolumn{2}{l}{ $P6_{3}/mcm$ ($\overline{1}$)}                                                                            \\
                                                 & (Ca,Ce)MnO$_{3}$ \cite{Vistoli2018}                                  & 20 nm   & 1               &                  & 15 (1-1.5)         & \multicolumn{2}{l}{$Pnma$ ($Pn'ma' $)}                                                                          \\
                                                 & CoNb$_{3}$S$_{6}$* \cite{Ghimire2018}         & SC& 1               & 27               & 23 (0)           & $P6_{3}22 $                             &                                                                        \\
\multirow{-5}{*}{CAF}                      & CoNb$_{3}$S$_{6}$*   \cite{Tenasini2020}      & 40-90 nm   & 2               & 400              & 5 (0)         & $P6_{3}22$                              &                                                                        \\ \hline
                                                 & Pr$_{2}$Ir$_{2}$O$_{7}$*   \cite{Machida2010,Ohtsuki2019}    &     & 1               & 10               & 1 (0)          &  $Fd\overline{3}m$ & UCu$_{5}$ \cite{Ueland2012}, pyrochlore\cite{Kim2020b}                                \\
\multirow{-2}{*}{NCP}                    & Mn$_{5}$Si$_{3}$ \cite{Surgers2014}           & SC   & 2               & 102              & 25 (5)           & $P6_{3}/mcm$ ($PCbcn$)                    & 40-160 nm \cite{Surgers2016,Surgers2017} \\ \hline
                                                 & \textit{GdPtBi} \cite{Suzuki2016}                                              & SC    & 60              & 30-200           & 10 (4)           & $F\overline{4}3m$ ($C_{c}c$)                        & Gd, NdPtBi \cite{Shekhar2018a}                     \\
\multirow{-2}{*}{C}       & \textit{EuTiO$_{3}$}   \cite{Takahashi2018}    &      & 5               & 20               & 2 (2)             & $I4/mcm$ ($Fm'mm$)                     & (Eu,Sm)TiO$_{3}$, \cite{Ahadi2018b} pyrochlore\cite{Ueda2018}                       

\end{tabular}
\caption{List of material representations of  anomalous Hall antiferromagnets. We group the materials into four archetypes: non-collinear (NCAF), collinear (CAF), non-coplanar (NCP) and canted (C). We also list whether the material was synthesized in bulk single-crystal (SC) or thin film (of a given thinckness). Next we list the  Hall resistivity and conductivity,  temperature of the experiments and applied magnetic field. Finally we give the crystal space group (SG) and magnetic space group (MSG).  Star marks systems with unresolved magnetic ordering.}
\end{table}

\twocolumngrid

\clearpage

\subsection*{Perspectives}

In this last section we look at  anomalous Hall antiferromagnets from a broader perspective. 
We discuss their emerging role at the intersection of fundamental and applied physics fields involving multipole magnetism, spin-momentum locking, topological phases, 
spintronics, and dissipationless nano-electronics.

%

Louis N\'eel stated in his Nobel lecture that antiferromagnetic substances do not appear to have any practical applications.\cite{Neel1971} 
In another part of N\'eel's lecture, he mentioned that effects depending on the square of the 
spontaneous (sublattice) magnetization should show the same variation in antiferromagnets as in ferromagnets.\cite{Neel1971}
Initial work in the  nascent field of antiferromagnetic spintronics 
discounted N\'eel's first statement, and  invoked his second 
by appealing to magnetic anisotropy energy for memory, anisotropic magnetoresistance for reading, and, 
damping-like spin-torque for writing functionalities.\cite{MacDonald2011,Gomonay2014,Jungwirth2016} 
Consistent with N\'eel's second statement, the magnetic anisotropy energy is given by terms in the thermodynamic potential that
 are even in the sublattice magnetization ${\bf M}_i$,\cite{Neel1971,Landau1984} and the anisotropic magnetoresistance corresponds to the ${\cal T}$-even, and 
 therefore also ${\bf M}_i$-even, symmetric conductivity components.\cite{Shick2010,Park2011b,Marti2014,Kriegner2016,Jungwirth2016}
 The damping-like torque $\sim {\bf M}_i\times({\bf M_i}\times{\bf p})$ is in the same category.\cite{MacDonald2011,Gomonay2014,Jungwirth2016}
 Here  {\bf p} is a uniformly spin-polarized current which can be injected into the antiferromagnet from a relativistic injector.\cite{Zelezny2014} 
 No ferromagnet is needed for any of these three basic functionalities of a relativistic antiferromagnetic spintronic memory device.\cite{Jungwirth2016} 

These initial efforts \cite{Turov1965,Neel1971,Nunez2006,Surgers2014,Surgers2016,Ghimire2018} did not consider
 leveraging favorable symmetry types, 
 and were instead firmly embedded within a traditional picture focusing just on the mutually compensating role of spins on magnetic sublattices.  
Antiferromagnetic spintronics made a step forward with the demonstration of electrical switching in a lower-symmetry collinear two-sublattice antiferromagnet
by a field-like spin-torque $\sim {\bf M_i}\times{\bf p}_i$, where  ${\bf p}_i$ has opposite sign on the two sublattices. This 
enabled the first experimental demonstration of functional antiferromagnetic memory cells.\cite{Zelezny2014,Wadley2016,Jungwirth2016,Wadley2018,Bodnar2018,Olejnik2017,Zelezny2018} 
The devices in question employed the antiferromagnetic crystals CuMnAs or Mn$_2$Au, both of which have ${\cal T}$-symmetry and ${\cal P}$-symmetry broken, but are ${\cal PT}$-invariant.\cite{Zelezny2014,Wadley2016} As we have explained, this still protects the  spin-degeneracy of the energy bands across the entire Brillouin zone and corresponds to ${\cal T}$-symmetric Laue groups, i.e., excludes the magnetic dipole and the ${\cal T}$-odd linear response Hall effect. However, the ${\cal PT}$-invariance accompanied by broken ${\cal P}$ and ${\cal T}$ symmetries, and the related (polar) magnetic toroidal  dipole (Fig.~2b),\cite{Watanabe2018a,Hayami2018,Thole2020} signal richer physics than anticipated within the traditional N\'eel picture of antiferromagnets. For example, besides being switchable by a field-like spin-torque,\cite{Zelezny2014,Wadley2016,Jungwirth2016,Bodnar2018,Zelezny2018,Watanabe2018a} the magnetic toroidal  dipole antiferromagnets also allow for electrical detection of the N\'eel vector reversal by a ${\cal T}$-odd second-order magnetoresistance.\cite{Watanabe2018a,Godinho2018} 

The spintronic phenomena discussed above are typically considered in systems with an ordinary metallic conduction. 
Antiferromagnets, however, occupy a rich materials landscape which makes it possible to realize more exotic phases associated with the ${\cal PT}$-symmetry. 
These include antiferromagnetic Dirac semimetals with a topological metal-insulator transition\cite{Smejkal2016,Tang2016} and magneto-electric axion topological antiferromagnets.\cite{Mogi2017,Grauer2017,Xiao2018,Marsh2019}

The discovery of a robust room-temperature  Hall effect in compensated non-collinear Mn$_3$X (X=Ir,Sn,Ge,Pt) magnets\cite{Chen2014,Kubler2014,Nakatsuji2015,Kiyohara2015,Nayak2016,Liu2018b} was a breakthrough which went beyond N\'eel's  assumption of equivalence in  the behavior of antiferromagnets and ferromagnets only in effects that depend on the square of the spontaneous (sublattice) magnetization. Anomalous Hall antiferromagnets exhibit analogous behavior to ferromagnets for an effect which is odd in the (sublattice) magnetization. Remarkably, unlike ferromagnets, they exhibit the ${\cal T}$-odd Hall effect while having zero magnetic dipole in the absence of relativistic spin-orbit cpupling. On the other hand, the non-collinear magnetic order does generate a magnetic octupole\cite{Suzuki2017,Tsai2020,Nomoto2020} (Fig.~2d), which is the next term after the dipole in the ${\cal T}$-odd magnetic multipole expansion. 

Beside providing a mechanism for electrical readout functionality in  spintronic devices,\cite{Liu2018b,Tsai2020} the  Hall effect in the compensated Mn$_3$X magnets led to the realization of its optical and thermo-electric ${\cal T}$-odd counterparts,\cite{Sivadas2016,Ikhlas2017,Higo2018c,Feng2020b} known under the terms Kerr/Faraday effect and Nernst effect, respectively.\cite{Nemec2018,Mizuguchi2019} Moreover, prompted by the Hall effect studies, the compensated non-collinear magnets were also identified as generators of the spin-polarized currents,\cite{Zelezny2017a,Kimata2019a} which in ferromagnets underpin commercial electrical reading and writing based on giant (tunneling) magnetoresistance and spin-transfer torques.\cite{Chappert2007,Ralph2008,Brataas2012,Bhatti2017,Duine2018} Bipolar electrical current manipulation of compensated non-collinear magnets 
using the same protocol as developed for ferromagnetic spintronics has been demonstrated up to THz speeds.\cite{Tsai2020,Tsai2021,Miwa2021}

While replicating several of the key effects driving ferromagnetic spintronics, non-collinear magnetic order prohibits a generation of spin-conserved phenomena even when the relativistic spin-orbit coupling is diminished. Protecting the spin is, however, a central problem in spintronics in its quest to complement charge-based microelectronics.

The discovery of the Hall effect generated by the compensated
collinear magnetic order \cite{Smejkal2020,Feng2020a}  in  RuO$_2$ thus not only gave us the opportunity
 to explain the physics of anomalous Hall antiferromagnets in a pedagogical way, borrowing frequently from text-book symmetry formalisms developed for collinear magnets.\cite{Landau1984,Turov1965}  Anomalous Hall antiferromagnets
 represent an emerging class of materials, one that may allow us to combine the strengths of antiferromagnetic and ferromagnetic spintronics concepts and materials, 
 while  compensating for their respective weaknesses. In particular, experimental antiferromagnetic spintronic devices have already 
 showcased the utility of materials ranging from insulators and semiconductors to metals, while demonstrating insensitivity to magnetic field perturbations, 
and exploiting the greater flexibility in device-geometry allowed in the absence of the dipolar shape anisotropy.
They have also demonstrated electrical and optical writing pulse-lengths from millisecond to femtosecond, analog time-dependent logic-in-memory functionalities 
reminiscent of neuromorphic computing elements, and information coding into metastable atomic-scale magnetic textures.\cite{Jungwirth2016,Zelezny2018,Nemec2018,Gomonay2018,Baltz2018,Song2018b,Mizuguchi2019,Siddiqui2020,Fukami2020,Kurenkov2020,Kaspar2021,Krizek2020a,Higo2021} Ferromagnetic digital memories, on the other hand, owe their commercial success primarily to the giant magnetoresistive readout signals and efficient spin transfer torque writing, relying on spin-conserving electron transport.\cite{Chappert2007,Ralph2008,Brataas2012,Bhatti2017,Duine2018} 

The discovery of the Hall effect in the compensated collinear magnets has led directly to a theory proposal that the essential spintronic reading and writing principles based on conserved spin-currents should be readily available in  these systems.\cite{Reichlova2020,Gonzalez-Hernandez2021,Smejkal2021,Shao2021}  Here the suitable compensated collinear magnets have the non-relativistic alternating spin-polarization, and the corresponding Fermi surfaces can have a characteristic symmetry  of a magnetic toroidal quadrupole (Fig.~2c).
\cite{Smejkal2020,Lopez-Moreno2012,Noda2016,Ahn2019,Hayami2019,Yuan2020,Feng2020a,Reichlova2020,Hayami2020,Yuan2021a,Egorov2021,Smejkal2021,Gonzalez-Hernandez2021,Smejkal2021a} 

A recent study,\cite{Smejkal2021a} based on a non-relativistic spin-symmetry group formalism and focusing on collinear magnets, has established that the non-relativistic alternating spin-polarization in the momentum space  is not an exotic anomaly in some antiferromagnetic materials. Instead, it has classified the non-relativistic band structures of collinear magnets by three formally distinct and comparbly abundant spin-group types. They correspond, respectively, to (i) ferromagnets with the electronic structure split into majority and minority spin bands, (ii) antiferromagnets with spin-degenerate bands across the entire Brillouin zone, and (ii) the so called altermagnets whose spin polarization alternates in both real and momentum space while generating zero net magnetization.\cite{Smejkal2021a}

Finally, we highlight that  the Hall effect breakthroughs described in this review  open a new avenue in the research  of topological phases in condensed matter.\cite{Bradlyn2017,Vergniory2019,Elcoro2020,Xu2020} The DFT Hall effect studies of compensated non-collinear Mn$_3$X  or collinear CoNb$_3$S$_6$ magnets\cite{Felser2017,Kuroda2017,Yang2017c,Noky2019a,Smejkal2020,Chen2021a} draw the attention to the prospect of realizing Weyl semimetals with ${\cal T}$-symmetry breaking by the dipole-free magnetic order.  Similarly, the observed large room-temperature  Hall conductivities in the compensated non-collinear Mn$_3$Sn and Mn$_3$Ge and collinear RuO$_2$ magnetic crystals,\cite{Chen2014,Kubler2014,Nakatsuji2015,Kiyohara2015,Nayak2016,Smejkal2020,Feng2020a} which are metals but have a relatively small density of states at the Fermi level, open a new angle in the search for the quantized  Hall effect in  Chern insulators.\cite{Haldane1988,Dong2016,Zhou2016,Feng2020b} Moreover, the discovery of the strong alternating spin-splitting  in the non-relativistic band structure of the compensated collinear magnets\cite{Smejkal2020,Ahn2019,Hayami2019,Yuan2020,Feng2020a,Reichlova2020,Gonzalez-Hernandez2021} is a new opportunity for realizing the topological quantum phases at zero magnetic field, high temperatures, and in materials with abundant light elements. These are key prerequisites for  bring the dissipationless topological nano-electronics closer to practical applications.

%


%

\end{document}